\documentclass[prl,twocolumn]{revtex4}
\newcommand{\be}{\begin{equation}}
\newcommand{\ee}{\end{equation}}
\newcommand{\beqn}{\begin{eqnarray}}
\newcommand{\eeqn}{\end{eqnarray}}

\begin{document}

\title{{\bf Quantum Gravity and Lorentz invariance violation in
the Standard Model}}

\author{Jorge Alfaro}

\affiliation{Dept. de F\'\i sica Te\'orica C-XI, Facultad de Ciencias, Univ. Aut\'onoma
de Madrid, Cantoblanco, 28049, Madrid, Spain and \\Facultad de F\'{\i}sica, Pontificia Universidad Cat\'{o}lica de Chile \\
        Casilla 306, Santiago 22, Chile.\footnote{Permanent Address.}
\\ {\tt jalfaro@puc.cl}}

\date{\today}

\begin{abstract}
The most important problem of fundamental Physics is the quantization of the gravitational field.
A main difficulty is the lack of available experimental tests that discriminate among the theories
proposed to quantize gravity. Recently, Lorentz invariance violation by Quantum Gravity(QG) have been
the source of a growing interest. However, the predictions depend on ad-hoc hypothesis and too many
arbitrary parameters. Here we show that the Standard Model(SM) itself contains tiny Lorentz invariance
violation(LIV) terms coming from QG. All terms depend on one arbitrary parameter $\alpha$ that set the scale
of QG effects. This parameter can be estimated using data from the Ultra High Energy Cosmic Rays
spectrum to be $|\alpha|<\sim 10^{-22}-10^{-23}$.
\end{abstract}

\maketitle

 In recent years several proposal have been advanced to select theories and predict new phenomena
 associated to the Quantum gravitational field \cite{1,2,3,4}. Most of the new phenomenology is associated
 to some sort of Lorentz invariance violations(LIV's)\cite{5,6}. Recently \cite{7},
 this approach has been subjected to severe criticism.

In this letter, we assert that the main effect of QG is to deform the measure of integration of Feynman graphs
at large four momenta by a tiny LIV. The classical lagrangian is unchanged.
Equivalently, we can say that QG deforms the metric of space-time, introducing a tiny LIV
proportional to (d-4)$\alpha$, d being the dimension of space time in Dimensional Regularization
and $\alpha$ is the only arbitrary parameter in the model. Such small
LIV could be due to quantum fluctuations of the metric of space-time produced by QG:virtual black holes as
suggested in\cite{1}, D-branes as in \cite{e2}, compactification of extra-dimensions or spin-foam anisotropies
\cite{8}. A  precise derivation of $\alpha$ will have to wait for additional progress
in the available theories of QG\footnote{Such derivation must explain why the LIV
parameter is so small. Progress in this direction is in \cite{3,4,9}. There $\alpha$ appears as $(l_P/{ L})^2$,
where $l_P$ is Planck's lenght and ${ L}$ is defined by the semiclassical gravitational state
in Loop Quantum Gravity. If $L\sim 10^{11} l_P$, an $\alpha$ of the right order is obtained}

It is possible to have modified dispersion relations without a preferred frame(DSR)\cite{dsr}.
Notice, however, that in our case the classical lagrangian is invariant under usual linear Lorentz
transformations but not under DSR. So our LIV is more akin to radiative breaking of usual
Lorentz symmetry than to DSR. Moreover the regulator R defined below and the deformed metric (5) are given in a particular
inertial frame, where spatial rotational symmetry is preserved. That is why,
in this paper we are ascribing to the point of view of \cite{6} which is
widely used in the literature. The preferred frame  is the one where the Cosmic
Background Radiation is isotropic.

Within the Standard Model, such LIV implies several remarkable effects, which are wholly determined 
up to one arbitrary parameter ($\alpha$).The main effects are:

The maximal attainable velocity for
particles is not the speed of light, but depends on the specific couplings of the particles within 
the Standard Model. Noticeably, this LIV of the dispersion relations is the only acceptable,
according to the
very stringent bounds coming from the Ultra High Energy Cosmic Rays (UHECR) spectrum\cite{exp,9}.
Moreover, the specific interactions between particles in the SM,
determine different maximum attainable velocities  for each particle, a necessary requirement to explain
the Greisen\cite{10},Zatsepin and Kuz'min\cite{11}(GZK) anomaly\cite{6,9,old}. Since the Auger\cite{auger}
experiment is expected to produce results in the near future, powerful tests
of Lorentz
invariance using the spectrum of  UHECR will be available.

Also birrefringence occurs for charged leptons, but not for gauge bosons.
In particular, photons and neutrinos have different maximum attainable
velocities. This could be tested in the next generation of neutrino detectors such as
NUBE\cite{12,13}.

Vertices in the SM will pick up a finite LIV.

{\bf Cutoff regulator:}To see what are the implications of the asymmetry in the measure for renormalizable
theories, we will mimic the Lorentz asymmetry of the measure by the replacement
$$
\int d^dk->\int d^d k R(\frac{k^2+\alpha k_0^2}{\Lambda^2})
$$
Here R is an arbitrary function, $\Lambda$
is a cutoff with mass dimensions, that will go to infinity at the end of the calculation.
We normalize $R(0)=1$ to recover the original integral. $R(\infty)=0$ to regulate the integral.
$\alpha$ is a real parameter. Notice that we are assuming that rotational invariance in space is preserved.
More general possibilities such as violation of rotational symmetry in space can be easily incorporated
in our formalism.

This regulator has the property that for logarithmically divergent integrals, the divergent term is
Lorentz invariant whereas when the cutoff goes to infinity a finite LIV part proportional
to $\alpha$ remains.

{\bf One loop} Let D be the naive degree of divergence of a One Particle Irreducible (1PI) graph.
The change in the measure induces
modifications to the primitively log divergent integrals(D=0) In this case, the correction
amounts to a finite
LIV. The finite part of 1PI Green functions will not be affected.
Therefore, Standard Model predictions are intact, except for the maximum attainable velocity
for particles\cite{6} and interaction vertices, which receive a finite wholly determined contribution from Quantum Gravity.

Let us analyze the primitivily divergent 1PI graphs for bosons first.

{\bf Self energy:}
$\chi(p)=\chi(0)+A^{\mu\nu}p_\mu p_\nu +convergent$,
$A^{\mu\nu}=\frac{1}{2}\partial_\mu\partial_\nu\chi(0)$. We have:
$$
A^{\mu\nu}=c_2\eta^{\mu\nu}+a^{\mu\nu}
$$
$c_2$ is the log divergent wave function renormalization counterterm; $a^{\mu\nu}$
is a finite LIV. The on-shell condition is:
$$
p^2-m^2-a^{\mu\nu}p_\mu p_\nu=0
$$
If spatial rotational invariance is preserved, the nonzero components of the matrix $a$ are:
$$
a^{00}=a_0;\ \
a^{ii}=-a_1
$$
So the maximum attainable velocity for this particle will be:
\begin{equation}
v_m=\sqrt{\frac{1-a_1}{1-a_0}}\sim 1-(a_1-a_0)/2
\end{equation}

For fermions, we have the self energy graph
$$
\Sigma(p)=\Sigma(0)+s^{\mu\nu}\gamma_\nu p_\mu
$$
$s^{\mu\nu}\gamma_\nu =\partial_\mu\Sigma(0)$. Moreover
$$
s^{\mu\nu}=s\eta^{\mu\nu}+a^{\mu\nu}/2
$$
$s$ is a log divergent wave function renormalization counterterm; $a^{\mu\nu}$
is a finite LIV. The maximum attainable velocity of this particle will be given again by equation (1).

By doing explicit computations for all particles in the SM, we get definite predictions
for the LIV,
assuming a particular regulator $R$. However, the dependence on $R$ amounts to a
multiplicative factor.
So ratios of LIV's are uniquely determined.

{\bf Vertex correction}
This graph has $D=0$, so the regulator R  will induce
a tiny LIV.

{\bf Gauge Bosons}
Consider the most general quadratic
Lagrangian which is gauge invariant, but could permit LIV's \footnote{A Chern-Simons
term is absent due to the symmetry $k_\mu->-k_\mu$, which is preserved by the regulator.}
$$
L=c^{\mu\nu\alpha\beta}F_{\mu\nu}F_{\alpha\beta}
$$
$c^{\mu\nu\alpha\beta}$ is antisymmetric in $\mu\nu$ and $\alpha\beta$ and symmetric by
$(\alpha,\beta)<->(\mu,\nu)$
It implies
that the most general expression for the self-energy of the gauge boson will be

\be
\Pi^{\nu\beta}(p)=c^{\mu\nu\alpha\beta}p_\alpha p_\mu \Pi(p)
\ee

We see that
$$
p_\nu \Pi^{\nu\beta}(p)=0
$$

$c^{\mu\nu\alpha\beta}$ is given by a logarithmically divergent integral.We get:
\be
c^{\mu\nu\alpha\beta}=c_2(\eta^{\mu\alpha}\eta^{\nu\beta}-\eta^{\mu\beta}\eta^{\nu\alpha})+
a^{\mu\nu\alpha\beta}
\ee
$c_2$ is a Lorentz invariant counterterm and $a^{\mu\nu\alpha\beta}$ is a LIV.

It is clear that
the same argument applies to massive gauge bosons that got their mass by spontaneous gauge
symmetry breaking
as well as to the graviton in linearized gravity.

Explicit computations are simplified by using Dimensional
Regularization as explained below.

{\bf LIV Dimensional Regularization}
We generalize dimensional regularization to a d dimensional space with an arbitrary constant
metric $g_{\mu\nu}$. We work with a positive definite metric first  and then
Wick rotate. We will illustrate the procedure with an example.
Here $g=det(g_{\mu\nu})$ and $\Delta>0$.

\beqn
\frac{1}{\sqrt{g}}\int\frac{d^dk}{(2\pi)^d}\frac{k_\mu k_\nu}{(k^2+\Delta)^n}=\nonumber\\
\frac{1}{\sqrt{g}\Gamma(n)}\int_0^\infty dt t^{n-1}\int\frac{d^dk}{(2\pi)^d}k_\mu k_\nu e^{-t(g^{\alpha\beta}k_\alpha k_\beta+\Delta)}=\nonumber\\
\frac{1}{(4\pi)^{d/2}}\frac{g_{\mu\nu}}{2}\frac{\Gamma(n-1-d/2)}{\Gamma(n)}\frac{1}{\Delta^{n-1-d/2}}
\eeqn

In the same manner, after Wick rotation, we obtain Appendix A4 of \cite{12}.

These definitions preserve gauge invariance, because the integration measure
is invariant under shifts.
To get a LIV measure, we assume that
\be
g^{\mu\nu}=\eta^{\mu\nu}+(4\pi)^2\alpha\eta^{\mu 0}\eta^{\nu 0}Res_{\epsilon=0}
\ee
where $\epsilon=2-\frac{d}{2}$ and $Res_{\epsilon=0}$ is the residue of the pole at $\epsilon=0$. A formerly divergent integral will have a pole at $\epsilon=0$, so
when we take the physical limit, $\epsilon->0$, the answer will contain a LIV term.

That is, LIV dimensional regularization consists in:

 1)Calculating the d-dimensional integrals using a general metric $g_{\mu\nu}$.

 2) Gamma matrix algebra is generalized to a general metric $g_{\mu\nu}$.

 3) At the end of the calculation, replace
 $g^{\mu\nu}=\eta^{\mu\nu}+(4\pi)^2\alpha\eta^{\mu 0}\eta^{\nu 0}Res_{\epsilon=0}$
 and then take the limit $\epsilon->0$.

To define the counterterms, we used the minimal substraction scheme(MSS); that is
we substract the poles in $\epsilon$ from the 1PI graphs.

As a concrete example, let us evaluate a typical one loop integral that appears in
the calculation of self energy graphs:
\beqn
A^{\mu\nu}=\int \frac{d^dk}{(2\pi)^d}\frac{k^\mu k^\nu}{[k^2-m^2+i0]^3}=\\
\frac{i}{(4\pi)^{d/2}}\frac{g^{\mu\nu}}{2}\frac{\Gamma(2-\frac{d}{2})}{2}\frac{1}{(m^2)^{2-\frac{d}{2}}}\\
=\frac{i}{(4\pi)^{d/2}}\frac{\eta^{\mu\nu}+(4\pi)^2\alpha\delta^\mu_0\delta^\nu_0Res_{\epsilon=0}}{2}\frac{\Gamma(2-\frac{d}{2})}{2}\frac{1}{(m^2)^{2-\frac{d}{2}}}\\
=\frac{i}{4(4\pi)^2}(\frac{\eta^{\mu\nu}}{\epsilon}+(4\pi)^2\alpha\delta^\mu_0\delta^\nu_0) +{\rm a\  finite\  LI \ term}
\eeqn

LIV Dimensional Regularization reinforces our claim that these tiny LIV's originates in
Quantum Gravity. In fact the sole change of the metric of space time is a correction of order
 $\epsilon$ to the Minkowsky metric and this is the source of the effects studied above.
Quantum Gravity is the strongest candidate to produce such effects because the gravitational
field is precisely the metric of space-time and tiny LIV modifications to the flat Minkowsky
metric may be produced by quantum fluctuations.

 Using data from the UHECR spectrum \cite{9}
(see also \cite{6}) we get the order of magnitude of the LIV:
$(a1-a0)/2\sim 10^{-22}-10^{-23}$. From the results listed below, we get $|\alpha|<\sim 10^{-22}-10^{-23}$.

{\bf Explicit One loop computations:} We follow \cite{14,pokorski}
and use LIV Dimensional Regularization.

{\bf Photons}
The LIV photon self-energy in the SM is:
\beqn
L\Pi^{\mu\nu}(q)=-
\frac{23}{3}e^2\alpha q_\alpha q_\beta\nonumber\\
(\eta^{\alpha\beta}\delta^\mu_0\delta^\nu_0+\eta^{\mu\nu}\delta^\alpha_0\delta^\beta_0-\eta^{\nu\beta}\delta^\mu_0\delta^\alpha_0-\eta^{\mu\alpha}\delta^\nu_0\delta^\beta_0)
\eeqn
It follows that
the maximal attainable velocity is
\be
v_{\gamma}=1-\frac{23}{6}e^2\alpha
\ee

We have included coupling to quarks and charged leptons as well as 3 generations and color.

{\bf Neutrinos:} The maximal attainable velocity is
\be
v_\nu=1-(3+tan^2\theta_w)  \frac{g^2 \alpha}{8}
\ee
In this scenario, we predict that neutrinos \cite{13} emitted simultaneously
 with photons in gamma ray bursts will not arrive simultaneously to Earth . The time delay during a flight
 from a source situated at a distance
 $D$ will be of the order of $(10^{-22}-10^{-23}) D/c\sim 10^{-5 } - 10^{-6}$ s, assuming $D=10^{10}$ light-years.
 No dependence of the time delay on the energy of high energy photons or neutrinos  should be 
 observed(contrast with \cite{1}). Photons will arrive earlier(later) if $\alpha<0$($\alpha>0$).
 These predictions could be tested in the next generation of neutrino detectors \cite{14}.

Using  $R_\xi$-gauges we have checked  that the LIV is gauge invariant.
The gauge parameter affects the Lorentz invariant part only.

{\bf Electron self-energy in the Weinberg-Salam model. Birrefringence:}

Define: $e_L=\frac{1-\gamma^5}{2}e$, $e_R=\frac{1+\gamma^5}{2}e$, where $e$ is the electron field. We get
\beqn
v_L=1-(\frac{g^2}{cos^2\theta_w}(sin^2\theta_w-1/2)^2 +e^2+g^2/2)\frac{\alpha}{2};\\
v_R=1-(e^2+\frac{g^2sin^4\theta_w}{cos^2\theta_w})\frac{\alpha}{2}
\eeqn
The difference in maximal speed for the left and right helicities is $\sim( 10^{-23} -10^{-24})$.

{\bf Higher order loops}
The Standard Model in the LIV background metric studied here is a renormalizable and unitary theory.

If the coupling constants are small as in the
Electroweak theory, the dominant LIV is the one loop contribution. This is true
also for QCD due to asymptotic freedom, but extrapolation to lower energies
is not simple due to hadronization.

We have computed the main effects of the LIV metric in the Standard Model but other extensions
of it could be considered as well.

Our results are generic: All particles will have a modified maximum attainable velocity and
birrefringence occurs for charged leptons, but not for gauge bosons, due to the chiral nature of
the Electroweak couplings.


\section*{Acknowledgements}
 The work of JA is partially supported by Fondecyt 1010967, Ecos-Conicyt C01E05 and Secretaria de Estado de Universidades e 
Investigaci\'on SAB2003-0238(Spain). He wants to thank A.A. Andrianov for several useful remarks; and interesting 
conversations with H.A. Morales-T\'ecotl, L.F.Urrutia, D. Sudarsky, C. Kounnas, C. Bachas, V. Kazakov, A. Bilal, 
M.B. Gavela, E. Alvarez  and A. Gonz\'alez-Arroyo. He acknowledges the hospitality of the Perimeter Institute and Ecole Normale
Superieure,Paris.


\end{document}